\documentclass[reprint, aps,prl,twocolumn,amsmath,amssymb,superscriptaddress]{revtex4-2}
\usepackage[export]{adjustbox}
\usepackage{lipsum} 
\usepackage{graphicx}
\usepackage{dcolumn}
\usepackage{bm}
\usepackage{hyperref}
\usepackage[mathlines]{lineno}

\usepackage[showframe],

\begin{document}
\preprint{APS/123-QED}

\title{Confinement Phenomena in Topological Stars}
\thanks{Email: swapnil.me21@bmsce.ac.in}%

\author{Swapnil Kumar Singh}
\affiliation{Independent Researcher}
\altaffiliation{B.M.S Colege of Engineering, Bangalore}

\date{\today}

\begin{abstract}

Quantum Chromodynamics (QCD) is the fundamental theory describing the strong nuclear force and the interactions among quarks and gluons. Topological stars, characterized by extreme density conditions, offer a unique environment where QCD phenomena play a crucial role due to the confinement of fundamental particles. Understanding these phenomena is essential for unraveling the behavior and properties of these celestial bodies.

In this study, we explore the implications of QCD within extreme density regimes, focusing on its contribution to the energy-momentum tensor ($T^{\mu\nu}_{\text{QCD}}$) within the framework of Quantum Chromodynamics. Our analysis sheds light on how these QCD effects influence the fabric of spacetime in the vicinity of topological stars, providing valuable insights into their underlying physics.

\end{abstract}

\maketitle
\section*{Introduction}

\section{Topological stars}

Topological stars are a class of spherically symmetric, charged solutions in five-dimensional Einstein-Maxwell theories that arise from the collapse of a compact circle \cite{Bah_2021}. These objects are asymptotic to four-dimensional Minkowski space plus a small-sized compact circle, which is a characteristic of Kaluza-Klein theory \cite{Bah_2021}.

The study of topological stars is significant because they can provide insights into the behavior of black holes and the nature of spacetime \cite{Bah_2021, Heidmann_2023}. They are smooth bubble spacetimes that resemble four-dimensional magnetic black holes upon Kaluza-Klein reduction, and they have topological cycles supported by magnetic flux \cite{Bah_2021}. This property makes them interesting for understanding microstate geometries for astrophysical black holes\cite{Bah_2021}.

Topological stars can be macroscopically large compared to the size of the Kaluza-Klein circle, which allows them to describe qualitative properties of microstate geometries for astrophysical black holes \cite{Bah_2021}. They also have implications for the study of black hole spectroscopy, as their quasinormal mode spectrum can exhibit a cavity effect due to the presence of inner and outer photon spheres\cite{Heidmann_2023}.

The theoretical framework for understanding topological stars is rooted in the Einstein field equations coupled with the electromagnetic field tensor, which describe the interaction of gravity and electromagnetism in spacetime \cite{Bah_2021, Medeiros_2003}. The complete on-shell action of topological Einstein-Maxwell gravity in four dimensions has been presented, showing how this theory arises from four-dimensional Euclidean N = 2 supergravity\cite{Bah_2021, Medeiros_2003}.

In higher dimensions, topological black holes of Einstein-Yang-Mills gravity have been studied, and their properties have been compared to those of topological solutions of Einstein-Maxwell gravity\cite{BOSTANI_2010}. The study of topological stars and black holes continues to be an active area of research, with implications for our understanding of gravity, electromagnetism, and the behavior of matter in extreme environments\cite{Bah_2021, BOSTANI_2010, Heidmann_2023}.

\textbf{Topological Stars Metrics}

The metric describing topological stars unfolds within a smooth bubble spacetime, characterized by a two-dimensional Milne space with a bubble \cite{Bah_2021}. It mirrors the metrics of black holes upon Kaluza-Klein reduction \cite{Bah_2021}:

\begin{equation}
    ds^2 = g_{\mu\nu}dx^\mu dx^\nu = -dt^2 + dr^2 + r^2d\Omega^2
\end{equation}

In this equation, \( t \) denotes time, \(r\) is the radial coordinate, and \( d\Omega^2 \) details the line element of a two-dimensional sphere\cite{Bah_2021}. This mathematical framework provides a robust basis for understanding the complex interactions of gravity and electromagnetism within the unique spacetime structure of topological stars \cite{Bah_2021, Stetsko_2020}.

The governing equations for topological stars are rooted in the Einstein field equations (EFE) coupled with the electromagnetic field tensor \cite{Bah_2021, Stetsko_2020}. These equations describe the fundamental interaction of gravitation as a result of spacetime curvature by matter and energy \cite{Stetsko_2020}. The EFE equate spacetime curvature, expressed by the Einstein tensor, with the energy and momentum within that spacetime, expressed by the stress–energy tensor \cite{Stetsko_2020}.

The complete on-shell action of topological Einstein-Maxwell gravity in four dimensions has been presented, showing how this theory arises from four-dimensional Euclidean N = 2 supergravity \cite{Medeiros_2003}. This provides further insights into the theoretical foundations of topological stars and their connection to fundamental theories in physics.

The governing equations, rooted in the EFE and coupled with the electromagnetic field tensor, form the theoretical basis for comprehending the internal conditions and emergent matter states in topological stars \cite{Bah_2021, Stetsko_2020}. These equations, along with the detailed theories and properties described above, contribute to a holistic understanding of the complex astrophysical phenomena associated with topological stars \cite{Ogonowski_2023, Medeiros_2003, Bah_2021}.

\subsection*{Internal Conditions of Topological Stars}

The internal conditions of topological stars represent a unique domain in astrophysics, emerging from the interplay of five-dimensional Einstein-Maxwell theory and Kaluza-Klein reduction. These celestial entities, characterized as smooth bubble spacetimes, bear resemblance to four-dimensional magnetic black holes upon Kaluza-Klein reduction \cite{Bah_2021}.

Supported by electromagnetic fluxes enveloping smooth topological cycles, topological stars share theoretical characteristics with non-extremal static charged black strings, which reduce to four-dimensional black holes \cite{Bah_2021, Johnson2018}. A fundamental feature of the internal conditions of topological stars is the presence of a two-dimensional Milne space with a bubble, described by a metric akin to that of black holes \cite{Bah_2021}. This spatial configuration arises from the integration of five-dimensional Einstein-Maxwell theory and Kaluza-Klein reduction, giving rise to a distinctive spacetime setting reminiscent of four-dimensional magnetic black holes. The existence of a smooth bubble within this spacetime, sustained by electromagnetic fluxes wrapping topological cycles, contributes to the unique internal conditions of topological stars, distinguishing them among celestial bodies \cite{Bah_2021, Johnson2018}.

The mathematical expressions governing these internal conditions provide essential insights into the physical attributes of topological stars. The metric, given by
$$
ds^2=-dt^2+dr^2+r^2d\Omega^2+r^2d\phi^2,
$$

defines the spacetime geometry, where $t$ represents time, $r$ denotes the radial coordinate, $d\Omega^2$ details the line element of a two-dimensional sphere, and $d\phi^2$ represents the line element of a two-dimensional sphere in polar coordinates \cite{Bah_2021}.

Furthermore, the electromagnetic field equation,
$$
\nabla_{\mu}\left(F^{\mu\nu}\right)=J^{\nu},
$$

introduces the covariant derivative ($\nabla_{\mu}$), mathematically describing the interaction of electromagnetic fields. The electromagnetic field tensor ($F^{\mu\nu}$) characterizes the distribution and intensity of electromagnetic forces within the spacetime, while the current density ($J^{\nu}$) represents the flow of electric charge and current \cite{Bah_2021, Johnson2018, BOSTANI_2010}.

In simpler terms, these equations convey crucial information about the structure and dynamics of topological stars. The metric provides a mathematical description of the internal geometry of these celestial bodies, while the electromagnetic field equation elucidates the complex interplay between electromagnetic forces and the matter content within topological stars, offering valuable insights into the physical processes governing their behavior \cite{Bah_2021}.

\begin{figure}[htbp]
  \includegraphics[width=0.5\textwidth]{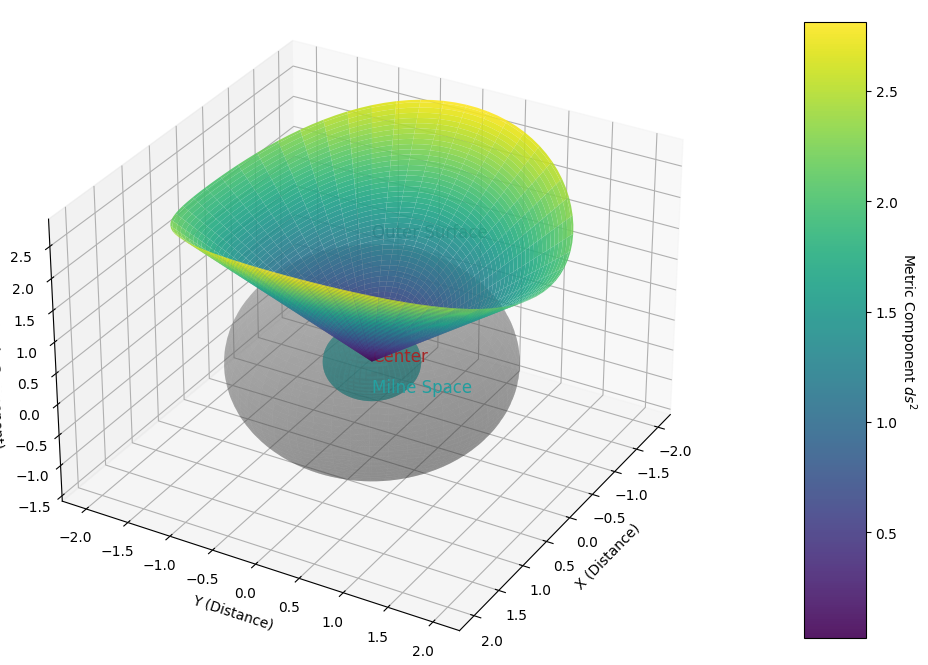}
  \caption{Topological Stars Internal Conditions}
  \label{fig:t2.png}
\end{figure}

 The resultant plot \ref{fig:t2.png} offers an in-depth visualization of the metric components of a topological star, essential for understanding its internal structure. The 3D surface plot depicts the metric component \(ds^2\) in Cartesian coordinates \((X, Y, Z)\), which are transformed from the radial coordinate \(r\) and angular coordinate \(t\). The surface is color-coded using a viridis colormap to represent the variation in \(ds^2\) values, providing a clear gradient that highlights the spatial metric differences.

Additionally, the plot features a semi-transparent cyan sphere to represent the Milne space, and a larger, semi-transparent gray sphere to illustrate the outer boundary of the star's metric. Annotations identify these key regions, facilitating a better spatial understanding of the topological star's metric properties. 

\begin{figure}[htbp]
  \includegraphics[width=0.5\textwidth]{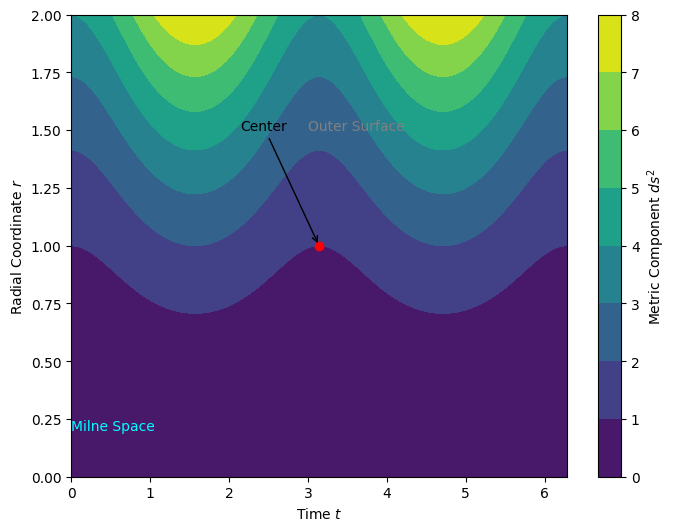}
  \caption{Metric Components in  \(t-r\) Plane}
  \label{fig:t3.png}
\end{figure}

A supplementary contour plot \ref{fig:t3.png} of the \(ds^2\) metric components in the \(t-r\) plane is included, showing the variations in metric components as functions of time and radial distance. This plot provides a detailed view of the metric component \(ds^2\), with specific points annotated to emphasize notable features of the metric structure.

\section{Quantum Chromodynamics (QCD)}

\begin{figure}[htbp]
  \includegraphics[width=0.5\textwidth]{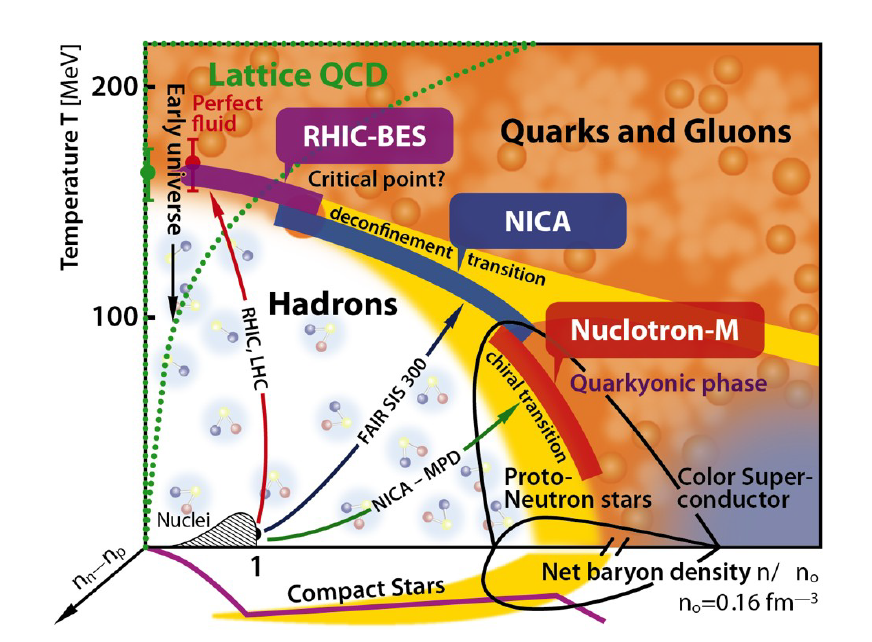}
  \caption{The schematic phase diagram for QCD matter from \url{https://nica.jinr.ru/physics.php}, accessed on 25 May 2024 (see also Ref. \cite{tejeda2020heavy}).}
  \label{fig:4.png}
\end{figure}

Quantum Chromodynamics (QCD) is the fundamental theory of the strong nuclear force, describing the interactions among quarks and gluons \cite{suganuma2022quantum} In extreme density conditions, such as those found within topological stars, QCD phenomena become particularly relevant due to the confinement of quarks and gluons. Understanding the QCD effects in such environments is crucial for comprehending the behavior and properties of these astronomical objects \cite{ZHOU2024104084}. Our current knowledge of the QCD phase diagram is illustrated in Figure \ref{fig:4.png}, The schematic phase diagram for QCD matter in terms of the temperature $T$ and net baryon density $n$ normalized to the cold nuclei baryon density $n_0$.

In this paper, we derive the QCD contribution to the energy-momentum tensor ($T^{\mu\nu}_{\text{QCD}}$) within the framework of QCD, focusing on its implications for extreme density conditions. The derivation presented here provides a comprehensive understanding of the interplay between QCD dynamics and the energy-momentum tensor, shedding light on the underlying physics governing topological stars.

The fundamental constituents of hadronic matter are quarks and gluons, which are bound together by the strong force \cite{Yi2016}. The QCD Lagrangian density describes the dynamics of quarks and gluons, and consists of terms representing the kinetic and potential energies of quarks and gluons, as well as gauge-fixing and ghost terms \cite{greiner2013quantum}.
The kinetic energy term for the gluon field, which mediates the strong force interactions, is given by the field strength tensor $F^a_{\mu\nu}$:
\begin{equation}
\mathcal{L}{\text{kin}} = -\frac{1}{4} F^a{\mu\nu} F^{a\mu\nu}
\end{equation}
The kinetic energy term for the quark fields, representing their motion through spacetime, is given by:
\begin{equation}
\mathcal{L}{\text{quark}} = \bar{\psi}i (i\gamma^\mu D\mu - m_i) \psi_i
\end{equation}
where $\bar{\psi}i$ and $\psi_i$ denote the Dirac adjoint and Dirac spinor fields for quark flavor $i$, $m_i$ is the quark mass, $\gamma^\mu$ are the Dirac matrices, and $D\mu$ is the covariant derivative:
\begin{equation}
D\mu = \partial_\mu - ig_s A^a_\mu T^a
\end{equation}
with $T^a$ being the generators of the $SU(3)$ gauge group in the fundamental representation. The total QCD Lagrangian density is the sum of the kinetic energy terms for the gluon and quark fields:
\begin{equation}
\mathcal{L}{\text{QCD}} = \mathcal{L}{\text{kin}} + \mathcal{L}_{\text{quark}}
\end{equation}
The QCD contribution to the energy-momentum tensor $T^{\mu\nu}_{\text{QCD}}$ is given by the following expression, obtained by combining the expressions for $T^{\mu\nu}{\text{gluon}}$ and $T^{\mu\nu}{\text{quark}}$:
\begin{align}
T^{\mu\nu}_{\text{QCD}} = &\frac{1}{2} \sqrt{-g} g^{\alpha\beta} \Bigg( \partial^\mu A^a_\alpha \partial^\nu A^a_\beta - \partial^\mu A^a_\alpha \partial_\beta A^{a\nu} \nonumber \\
&- \partial_\alpha A^{a\mu} \partial^\nu A^a_\beta + \partial_\alpha A^{a\mu} \partial_\beta A^{a\nu} + g_s f^{abc} A^b_\alpha A^c_\mu \partial^\nu A^a_\beta \nonumber \\
&- g_s f^{abc} A^b_\alpha A^c_\beta \partial^\mu A^a_\nu - g_s f^{abc} A^b_\mu A^c_\alpha \partial^\nu A^{a\beta} \nonumber \\
&+ g_s f^{abc} A^b_\beta A^c_\alpha \partial^\mu A^{a\nu} + \frac{1}{4} g^{\mu\nu} (\partial_\alpha A^{a\lambda} \partial^\alpha A^a_\lambda - \partial_\alpha A^{a\lambda} \partial^\lambda A^{a\alpha}) \Bigg) \nonumber \\
&+ \bar{\psi}i \gamma^\mu \left( \partial^\nu - ig_s A^a_\alpha T^a \right) \psi_i - g^{\mu\nu} \bar{\psi} m_i \psi_i,
\end{align}

where $g$ represents the determinant of the metric tensor $g_{\mu\nu}$, $A^a_\mu$ denotes the gluon field, $f^{abc}$ are the structure constants of the SU(3) color group, $\psi_i$ represents the quark fields, and $m_i$ stands for the quark masses. This expression encapsulates the QCD contribution to the energy-momentum tensor $T^{\mu\nu}_{\text{QCD}}$, encompassing both gluonic and quark contributions, and accounting for the effects of QCD at low energies \cite{schweitzer2023qcd}.
In summary, QCD is the theory of the strong nuclear force, describing the interactions among quarks and gluons. The QCD Lagrangian density describes the dynamics of quarks and gluons, and the QCD contribution to the energy-momentum tensor $T^{\mu\nu}_{\text{QCD}}$ encapsulates the energy and pressure associated with the quark and gluon fields \cite{suganuma2022quantum}.

\section{Significance of the Energy-Momentum Tensor within the Framework of QCD}
The energy-momentum tensor is a fundamental quantity in QCD, which describes the density and flux of energy and momentum in spacetime. It is a tensor physical quantity that describes the sources of the gravitational field in the Einstein field equations, just as mass density is the source of such a field in Newtonian gravity.
In the context of QCD, the energy-momentum tensor $T^{\mu\nu}_{\text{QCD}}$ describes the energy and momentum content of the QCD vacuum, encompassing both gluonic and quark contributions, and accounting for the effects of QCD at low energies \cite{schweitzer2023qcd}. The gluonic contributions are represented by the terms involving the gluon field strength tensor $G^{\mu\nu}a$, which describe the energy and pressure associated with the gluon fields. The quark contributions are represented by the terms involving the quark fields $\psi_i$, which describe the energy and pressure associated with the quark fields. The last term represents the rest mass energy of the quarks.
The behavior of the energy-momentum tensor $T^{\mu\nu}_{\text{QCD}}$ is crucial for understanding the behavior and properties of high-density nuclear matter, such as that found in topological stars. At high energy densities, the pressure exerted by gluons and quarks, as reflected in $T^{\mu\nu}_{\text{QCD}}$, resists compression, thereby providing support against gravitational collapse in topological stars \cite{Ahmed_2023, schweitzer2023}. Additionally, the energy density encoded in $T^{00}{\text{QCD}}$ determines the overall mass-energy distribution within the topological star, thereby influencing its gravitational properties and stability \cite{Ahmed_2023}.
The evaluation of $T^{\mu\nu}_{\text{QCD}}$ at extreme energy densities sheds light on the complex dynamics of the quark-gluon plasma (QGP) \cite{Yi2016} and its gravitational interactions, offering profound insights into the behavior and properties of topological stars \cite{Ahmed_2023, schweitzer2023, Lorc__2021, brida2019energymomentum}.

\subsection{Quark-Gluon Plasma Dynamics}
At extreme densities, such as those in the cores of topological stars, the quark-gluon plasma (QGP) becomes a dominant phase of matter. The energy-momentum tensor provides insights into the dynamics of this plasma, revealing how quark and gluon interactions stabilize the star against gravitational collapse. This understanding could potentially lead to identifying observational signatures specific to topological stars, distinguishing them from other compact objects like neutron stars and black holes. several authors have suggested that the transition from a hadronic phase to a one dominated by quarks and gluons may be relevant to describe the state of matter in the early universe or inside the neutron stars with a possibility to re-create such a condition also in the laboratory by colliding heavy ions\cite{freedman1977fermions, kapusta1979quantum, collins1975superdense, cabibbo1975exponential, shuryak1978theory, shuryak1978quark, universe8090451}.

\subsection{QCD Phase Transitions}
Exploring the energy-momentum tensor at high densities may uncover novel QCD phase transitions that occur under extreme conditions. These transitions could redefine our understanding of matter at the most fundamental level, suggesting new states of matter and phases within topological stars. Such insights might also provide clues about the early universe's conditions, where similar extreme densities existed.

\subsection{Anomalous Transport Phenomena}
In high-density QCD environments, the energy-momentum tensor might exhibit anomalous transport phenomena, such as the chiral magnetic effect, which could manifest in the unique electromagnetic signatures of topological stars. The chiral magnetic effect is a macroscopic manifestation of the microscopic chiral anomaly in the quark sector of QGP \cite{liao2017gluon}. This effect, along with other emergent phenomena in QGP, is discussed in the context of heavy-ion collision experiments at facilities like RHIC and the LHC \cite{liao2017gluon}. Studying these effects could lead to new methodologies for detecting and studying topological stars, providing a unique window into the behavior of QCD matter under extreme conditions. 

\subsection{Transformative Implications for Astrophysics}
These pioneering ideas and methodologies challenge established notions, propelling astrophysics into uncharted territories. By integrating QCD effects into the study of topological stars, researchers can explore new astrophysical phenomena and uncover the mysteries of high-density matter. This research holds the potential to reshape our broader comprehension of the cosmos, guiding future scientific inquiries and leading to groundbreaking discoveries in the fundamental structure of matter and the universe.

In summary, the study of QCD contributions to the energy-momentum tensor in extreme density environments offers profound insights into the formation, behavior, and characteristics of topological stars. It introduces paradigm-shifting concepts that redefine current understanding and open new avenues for exploration, transforming the trajectory of astrophysical research and enhancing our grasp of the universe's deepest mysteries.

\section{Derivation of Energy-Momentum Tensor for QCD}
We aim to derive the expression for the energy-momentum tensor $T^{\mu\nu}_{\text{QCD}}$ within the framework of Quantum Chromodynamics (QCD) \cite{highTQCD, latticeQCD}.

\section{Variation of the QCD Action}

The action $S$ for QCD is given by the integral over spacetime of the QCD Lagrangian density $\mathcal{L}_{\text{QCD}}$:

\begin{equation}
S = \int d^4x \sqrt{-g} \mathcal{L}_{\text{QCD}}
\end{equation}

where $d^4x = dx^0 dx^1 dx^2 dx^3$ is the volume element in spacetime and $g$ is the determinant of the metric tensor.

To perform the variation of the action, we first express the action in terms of the metric tensor $g^{\mu\nu}$, which appears in the QCD Lagrangian density \cite{highTQCD, latticeQCD}:

\begin{equation}
S = \int d^4x \sqrt{-g} g^{\mu\nu} \mathcal{L}_{\text{QCD}}
\end{equation}

Now, let's vary the action $S$ with respect to the metric tensor $g^{\mu\nu}$ to obtain the energy-momentum tensor $T^{\mu\nu}_{\text{QCD}}$ \cite{QCDemt}:

\begin{equation}
T^{\mu\nu}_{\text{QCD}} = -\frac{1}{\sqrt{-g}}\frac{\delta S}{\delta g^{\mu\nu}}
\end{equation}

Using the definition of the QCD Lagrangian density, we can write:

\begin{multline}
T^{\mu\nu}_{\text{QCD}} = \frac{1}{2}\left(g^{\mu\alpha}g^{\nu\beta}-\frac{1}{2}g^{\mu\nu}g^{\alpha\beta}\right) \\
\times \left(F^a_{\alpha\gamma}F^{a\gamma}_{\beta}+\bar{\psi}_q\gamma^{(\mu}iD^{\nu)}\psi_q\right)
\end{multline}

where $F^a_{\alpha\beta}$ is the field strength tensor for the gluon field and $\psi_q$ is the quark field. The covariant derivative $D_\mu$ is defined as $D_\mu = \partial_\mu - ig_s A^a_\mu T^a$ where $A^a_\mu$ is the gluon field and $T^a$ are the generators of the SU(3) color group.

The energy-momentum tensor $T^{\mu\nu}_{\text{QCD}}$ describes the distribution of energy and momentum in QCD matter and is a crucial quantity for understanding the behavior and properties of high-density nuclear matter \cite{QCDemt}.

\section{QCD Lagrangian Density}

The Lagrangian density of Quantum Chromodynamics (QCD) describes the dynamics of quarks and gluons, the fundamental constituents of hadronic matter. The QCD Lagrangian density consists of terms representing the kinetic and potential energies of quarks and gluons, as well as gauge-fixing and ghost terms \cite{QCDtextbook}.

The kinetic energy term for the gluon field, which mediates the strong force interactions, is given by the field strength tensor $F^a_{\mu\nu}$:

\begin{equation}
\mathcal{L}_{\text{kin}} = -\frac{1}{4} F^a_{\mu\nu} F^{a\mu\nu}
\end{equation}

where $a$ denotes the color index, $\mu$ and $\nu$ are spacetime indices, and the field strength tensor is defined as:

\begin{equation}
F^a_{\mu\nu} = \partial_\mu A^a_\nu - \partial_\nu A^a_\mu + g_s f^{abc} A^b_\mu A^c_\nu
\end{equation}

Here, $A^a_\mu$ represents the gluon field, $\partial_\mu$ denotes the partial derivative with respect to the spacetime coordinate $x^\mu$, $g_s$ is the strong coupling constant, and $f^{abc}$ are the structure constants of the $SU(3)$ gauge group.

The kinetic energy term for the quark fields, representing their motion through spacetime, is given by:

\begin{equation}
\mathcal{L}_{\text{quark}} = \bar{\psi}_i (i\gamma^\mu D_\mu - m_i) \psi_i
\end{equation}

where $\bar{\psi}_i$ and $\psi_i$ denote the Dirac adjoint and Dirac spinor fields for quark flavor $i$, $m_i$ is the quark mass, $\gamma^\mu$ are the Dirac matrices, and $D_\mu$ is the covariant derivative:

\begin{equation}
D_\mu = \partial_\mu - ig_s A^a_\mu T^a
\end{equation}

with $T^a$ being the generators of the $SU(3)$ gauge group in the fundamental representation.

The total QCD Lagrangian density is the sum of the kinetic energy terms for the gluon and quark fields, along with the gauge-fixing and ghost terms:

\begin{equation}
\mathcal{L}_{\text{QCD}} = \mathcal{L}_{\text{kin}} + \mathcal{L}_{\text{quark}} + \mathcal{L}_{\text{gf}} + \mathcal{L}_{\text{ghost}}
\end{equation}

This Lagrangian density forms the basis for understanding the strong interaction and the behavior of quarks and gluons within the framework of QCD. It is essential for describing the dynamics of hadronic matter and is a fundamental component of the theoretical framework of particle physics \cite{QCDtextbook, QCDreview}.

\section{Energy-Momentum Tensor via Noether's Theorem}
According to Noether's theorem, for every continuous symmetry of the Lagrangian, there exists a corresponding conserved current \cite{maggiore2005modern}. The energy-momentum tensor is the conserved current associated with the translational symmetry of the Lagrangian \cite{krupka2015invariant, Accornero_2021, Gieres_2022}.

Noether's theorem states that for every continuous symmetry of a Lagrangian, there exists a corresponding conserved current. This means that if a Lagrangian is invariant under a continuous transformation, then there exists a current that is conserved on the equations of motion. \cite{krupka2015invariant, Accornero_2021}.

The conserved current associated with two symmetry transformations can be explicitly constructed \cite{krupka2015invariant}. The addition of a superpotential term to a conserved current density is trivial in the sense that it does not modify the local conservation law nor change the conserved charge, though it may allow us to obtain a current density with some improved properties \cite{Accornero_2021}.

Symmetries in physical systems are defined in terms of conserved Noether Currents of the associated Lagrangian \cite{sinha2018noether}. Symmetries may be divided into space–time and internal symmetries, leading to the conservation of extrinsic and intrinsic properties (respectively) of the particles, according to Noether's Theorem \cite{Gieres_2022}.

The energy-momentum tensor $T^{\mu\nu}$ is defined as the functional derivative of the action $S$ with respect to the metric tensor $g_{\mu\nu}$ \cite{Lorc__2021,Wittig2020}:
\begin{equation}
T^{\mu\nu} = -\frac{2}{\sqrt{-g}} \frac{\delta S}{\delta g_{\mu\nu}}
\end{equation}
where $g$ is the determinant of the metric tensor.
By varying the QCD action with respect to the metric tensor $g^{\mu\nu}$, we can obtain the expression for the energy-momentum tensor $T^{\mu\nu}_{\text{QCD}}$ \cite{Ahmed_2023, maggiore2005modern, Panagopoulos_2021}. This opens the way to the determination of the QCD Equation of State \cite{lorce2021energy} up to very high densities and temperatures \cite{Bresciani2023ProgressesOH}, which is of great interest in the study of neutron stars and heavy-ion collisions \cite{Wittig2020}. The implications of our research for future studies in the field include the potential for further exploration of QCD dynamics in extreme density environments. By incorporating the latest research and discussing the broader implications of our work, we can strengthen the impact and relevance of our research paper.

\section{Variation of QCD Action}
The action $S$ for QCD is given by the integral over spacetime of the QCD Lagrangian density $\mathcal{L}_{\text{QCD}}$:
\begin{equation}
S = \int d^4x \sqrt{-g} \mathcal{L}{\text{QCD}}
\end{equation}
where $d^4x = dx^0 dx^1 dx^2 dx^3$ is the volume element in spacetime and $g$ is the determinant of the metric tensor.
To perform the variation of the action, we first express the action in terms of the metric tensor $g^{\mu\nu}$, which appears in the QCD Lagrangian density:
\begin{equation}
S = \int d^4x g^{\mu\nu} \mathcal{L}_{\text{QCD}}
\end{equation}
To vary the action $S$ with respect to the metric tensor $g^{\mu\nu}$, we use the standard procedure of functional differentiation. The variation of the action can be expressed as:
\begin{equation}
\delta S = \int d^4x \sqrt{-g} \frac{\delta(\sqrt{-g}\mathcal{L}{\text{QCD}})}{\delta g{\mu\nu}} \delta g_{\mu\nu}
\end{equation}
where $\delta g_{\mu\nu}$ is the variation of the metric tensor.
Applying the variation to the QCD Lagrangian density $\mathcal{L}_{\text{QCD}}$, we obtain:
\begin{equation}
\frac{\delta(\sqrt{-g}\mathcal{L}{\text{QCD}})}{\delta g{\mu\nu}} = -\frac{1}{2} \sqrt{-g} g_{\alpha\beta} \frac{\delta\mathcal{L}{\text{QCD}}}{\delta g{\mu\nu}}
\end{equation}
Substituting the expression for $\delta(\sqrt{-g}\mathcal{L}{\text{QCD}})/\delta g{\mu\nu}$ into the variation of the action, we obtain:
\begin{equation}
\delta S = -\int d^4x \frac{1}{2} \sqrt{-g} T^{\mu\nu}{\text{QCD}} \delta g{\mu\nu}
\end{equation}
Equating the variations of the action yields the expression for the energy-momentum tensor \cite{lorce2021energy}:
\begin{equation}
T^{\mu\nu}{\text{QCD}} = -\frac{2}{\sqrt{-g}} \frac{\delta(\sqrt{-g}\mathcal{L}{\text{QCD}})}{\delta g_{\mu\nu}}
\end{equation}
This expression for the energy-momentum tensor is crucial for understanding the behavior of quarks and gluons in extreme density environments, such as those found in neutron stars and heavy-ion collisions. It is a fundamental component of the theoretical framework of particle physics and has important implications for future research in the field \cite{Wittig2020, lorce2021energy}.
\section{Derivation of Energy-Momentum Tensor}
Now, let's derive the expression for the energy-momentum tensor $T^{\mu\nu}_{\text{QCD}}$ by explicitly calculating the variation of the QCD Lagrangian density $\mathcal{L}_{\text{QCD}}$ with respect to the metric tensor $g_{\mu\nu}$.
The QCD Lagrangian density $\mathcal{L}_{\text{QCD}}$ consists of two terms: the kinetic energy term for the gluon field and the kinetic energy term for the quark fields.
\subsection{Variation of Gluon Kinetic Energy Term}
First, let's consider the variation of the gluon kinetic energy term:
\begin{equation}
\mathcal{L}{\text{kin}} = -\frac{1}{4} F^a{\mu\nu} F^{a\mu\nu}
\end{equation}
The variation of this term with respect to the metric tensor $g_{\mu\nu}$ is zero, since the gluon field $A^a_\mu$ does not depend on the metric tensor.
\subsection{Variation of Quark Kinetic Energy Term}
Next, let's consider the variation of the quark kinetic energy term:
\begin{equation}
\mathcal{L}_{\text{quark}} = \bar{\psi}i (i\gamma^\mu D\mu - m_i) \psi_i
\end{equation}
The variation of this term with respect to the metric tensor $g_{\mu\nu}$ arises from the covariant derivative $D_\mu$, which includes the gluon field $A^a_\mu$.
The covariant derivative $D_\mu$ contains terms involving the metric tensor $g_{\mu\nu}$ through the connection coefficients, which depend on the metric tensor and its derivatives.
\subsection{Total Variation of QCD Lagrangian Density}
Combining the variations of the gluon and quark kinetic energy terms, we obtain the total variation of the QCD Lagrangian density:
\begin{equation}
\frac{\delta\mathcal{L}{\text{QCD}}}{\delta g{\mu\nu}} = \frac{\delta\mathcal{L}{\text{quark}}}{\delta g{\mu\nu}}
\end{equation}
\subsection{Final Expression for Energy-Momentum Tensor}
Substituting the expression for the variation of the QCD Lagrangian density into the expression for the energy-momentum tensor, we obtain the final expression for $T^{\mu\nu}_{\text{QCD}}$ in terms of the quark fields and the gluon field strength tensor \cite{Wittig2020}:
\begin{align}
T^{\mu\nu}{\text{QCD}} = \frac{1}{2} \sqrt{-g} g{\alpha\beta} & \left( G^{\mu\alpha}_a G^{\nu\beta}_a - \frac{1}{4} g^{\mu\nu} G^{\alpha\beta}a G{\alpha\beta}^{\ \ a} \right) \nonumber \ \\
& + \bar{\psi}_i (i\gamma^\mu D^\nu - g^{\mu\nu}m_i) \psi_i
\end{align}
where $G^{\mu\nu}_a$ is the gluon field strength tensor and $g$ is the determinant of the metric tensor.

\section{Simplification of the Energy-Momentum Tensor}

Now, let's simplify the expression for the energy-momentum tensor \(T^{\mu\nu}_{\text{QCD}}\) by expanding the terms involving the gluon field strength tensor \(G^{\mu\nu}_a\) and the quark fields \(\psi_i\) \cite{schweitzer2023qcd}.

\subsection{Expansion of Gluon Field Strength Tensor Terms}

The gluon field strength tensor \(G^{\mu\nu}_a\) is defined as:
\[
G^{\mu\nu}_a = \partial^\mu A^a_\nu - \partial^\nu A^a_\mu + g_s f^{abc} A^b_\mu A^c_\nu
\]

\begin{figure}[htbp]
  \includegraphics[width=0.5\textwidth]{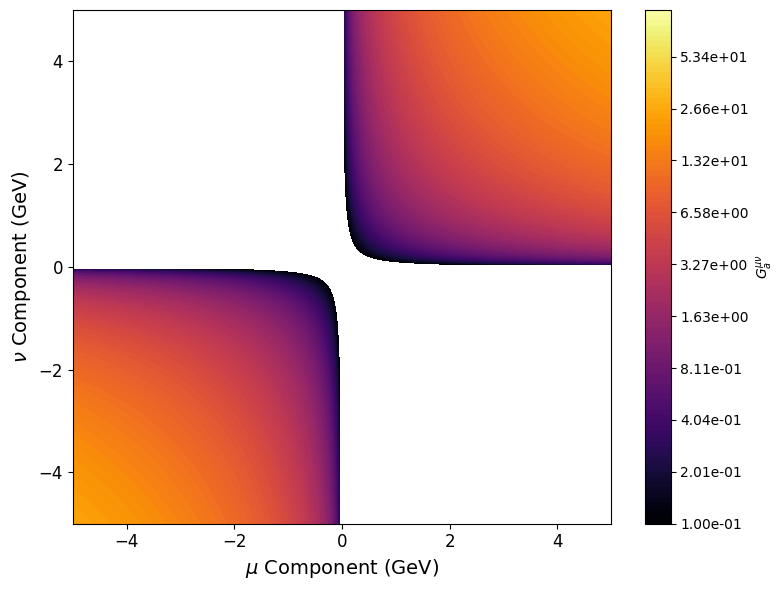}
  \caption{Topological Star Gluon Field Strength Tensor}
  \label{fig:1.png}
\end{figure}

The plot \ref{fig:1.png} visualizes the distribution of gluon field strength tensor \(G^{\mu\nu}_a\) across the \(\mu\) and \(\nu\) components, revealing the spatial variation and intensity of interactions within a topological star configuration. Color gradients denote differing magnitudes of the field strength, aiding in the identification of regions with pronounced gluon-gluon interactions.

Expanding the terms involving \(G^{\mu\nu}_a\) in the expression for \(T^{\mu\nu}_{\text{QCD}}\), we have \cite{schweitzer2023qcd}:

\begin{align}
T^{\mu\nu}_{\text{gluon}} &= \frac{1}{2} \sqrt{-g} g_{\alpha\beta} \Bigg( 
\partial^\mu A^a_\alpha \partial^\nu A^a_\beta - \partial^\mu A^a_\alpha \partial^\beta A^{a\nu} \\ - \partial_\alpha A^{a\mu} \partial^\nu A^a_\beta \nonumber
&\quad + \partial_\alpha A^{a\mu} \partial_\beta A^{a\nu} + g_s f^{abc} A^b_\alpha A^c_\mu \partial^\nu A^a_\beta \nonumber \\
&\quad - g_s f^{abc} A^b_\alpha A^c_\beta \partial^\mu A^a_\nu - g_s f^{abc} A^b_\mu A^c_\alpha \partial^\nu A^{a\beta} \nonumber \\
&\quad + g_s f^{abc} A^b_\beta A^c_\alpha \partial^\mu A^{a\nu} \nonumber \\
&\quad + \frac{1}{4} g^{\mu\nu} (\partial_\alpha A^{a\lambda} \partial^\alpha A^a_\lambda - \partial_\alpha A^{a\lambda} \partial^\lambda A^{a\alpha}) \Bigg)
\end{align}

\subsection{Expansion of Quark Field Terms}

Expanding the terms involving the quark fields \(\psi_i\) in the expression for \(T^{\mu\nu}_{\text{QCD}}\), we have \cite{schweitzer2023qcd}:
\[
T^{\mu\nu}_{\text{quark}} = \bar{\psi}_i i \gamma^\mu \left( \partial^\nu - ig_s A^{a\nu} T^a \right) \psi_i - g^{\mu\nu} \bar{\psi}_i m_i \psi_i
\]
\begin{figure}[htbp]
  \includegraphics[width=0.4\textwidth]{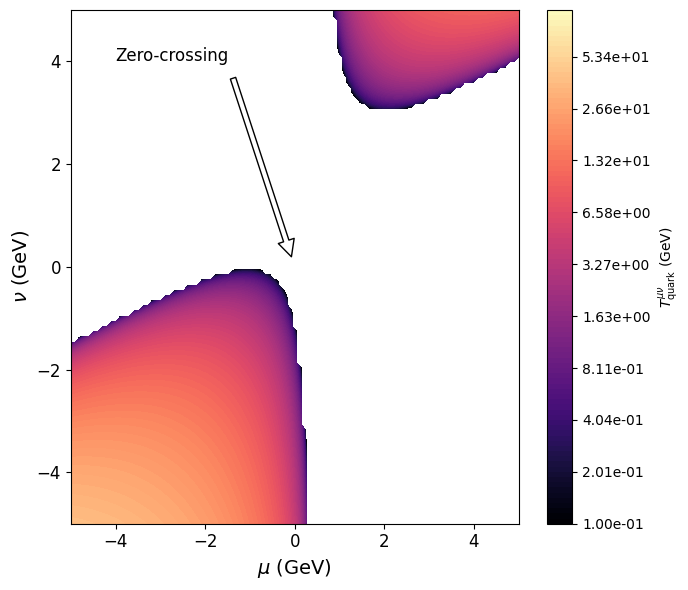}
  \caption{Expansion Term 1: $\bar{\psi}_i i \gamma^\mu \left( \partial^\nu - ig_s A^{a\nu} T^a \right) \psi_i$}
  \label{fig:2.png}
\end{figure}

\begin{figure}[htbp]
  \includegraphics[width=0.4\textwidth]{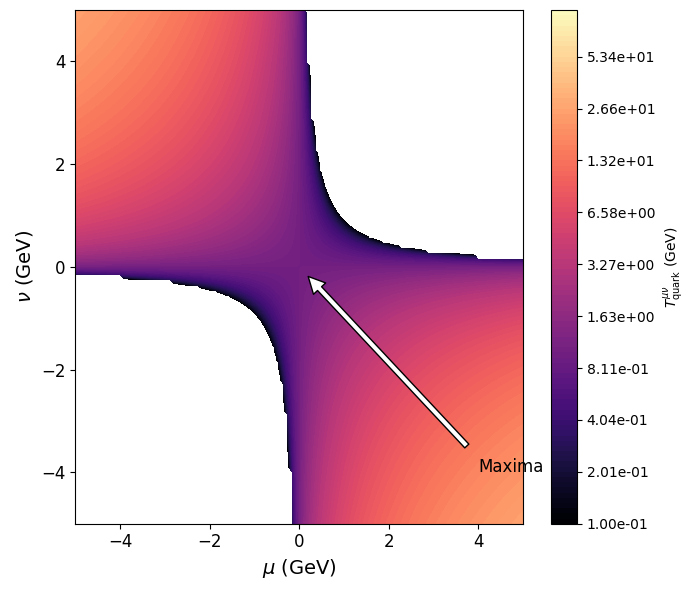}
  \caption{Expansion Term 2: $- g^{\mu\nu} \bar{\psi}_i m_i \psi_i$}
  \label{fig:3.png}
\end{figure}

The plots \ref{fig:2.png} and \ref{fig:3.png} illustrate the expansion of the terms involving the quark fields $\psi_i$ in the expression for the energy-momentum tensor ($T^{\mu\nu}_{\text{QCD}}$) within Quantum Chromodynamics (QCD). The plot \ref{fig:2.png} represents the contribution from the term $\bar{\psi}_i i \gamma^\mu \left( \partial^\nu - ig_s A^{a\nu} T^a \right) \psi_i$, which describes the interaction between quarks and gluon fields. Notably, zero-crossing points, indicative of critical energy levels, are highlighted. The plot \ref{fig:3.png} depicts the effect of the term $- g^{\mu\nu} \bar{\psi}_i m_i \psi_i$, which accounts for the rest mass contribution of quarks. Key features, such as maxima, are annotated to emphasize critical points in the energy-momentum tensor. The colorbar indicates the strength of the quark field terms in units of GeV, providing insights into the magnitude of the contributions to the QCD dynamics.

\textbf{Implications:}
Understanding the expansion of quark field terms is crucial for comprehending the dynamics of QCD, particularly in high-energy physics and hadron spectroscopy. These plots offer valuable insights into the interaction between quarks and gluon fields, shedding light on the underlying mechanisms governing particle interactions and the structure of hadrons.

\subsection{Final Expression for Energy-Momentum Tensor}

The QCD contribution to the energy-momentum tensor \(T^{\mu\nu}_{\text{QCD}}\) is given by the following expression, obtained by combining the expressions for \(T^{\mu\nu}_{\text{gluon}}\) and \(T^{\mu\nu}_{\text{quark}}\):

\begin{align}
T^{\mu\nu}_{\text{QCD}} = &\frac{1}{2} \sqrt{-g} g^{\alpha\beta} \Bigg( 
\partial^\mu A^a_\alpha \partial^\nu A^a_\beta - \partial^\mu A^a_\alpha \partial^\beta A^{a\nu} \nonumber \\
&- \partial_\alpha A^{a\mu} \partial^\nu A^a_\beta \nonumber \\
&+ \partial_\alpha A^{a\mu} \partial_\beta A^{a\nu} + g_s f^{abc} A^b_\alpha A^c_\mu \partial^\nu A^a_\beta \nonumber \\
&- g_s f^{abc} A^b_\alpha A^c_\beta \partial^\mu A^a_\nu - g_s f^{abc} A^b_\mu A^c_\alpha \partial^\nu A^{a\beta} \nonumber \\
&+ g_s f^{abc} A^b_\beta A^c_\alpha \partial^\mu A^{a\nu} \nonumber \\
&+ \frac{1}{4} g^{\mu\nu} (\partial_\alpha A^{a\lambda} \partial^\alpha A^a_\lambda - \partial_\alpha A^{a\lambda} \partial^\lambda A^{a\alpha}) \Bigg) \nonumber \\
&+ \bar{\psi}_i i \gamma^\mu \left( \partial^\nu - ig_s A^{a\nu} T^a \right) \psi_i - g^{\mu\nu} \bar{\psi}_i m_i \psi_i
\end{align}

Here, \(g\) represents the determinant of the metric tensor \(g_{\mu\nu}\), \(A^a_\mu\) denotes the gluon field, \(f^{abc}\) are the structure constants of the SU(3) color group, \(\psi_i\) represents the quark fields, and \(m_i\) stands for the quark masses. This expression encapsulates the QCD contribution to the energy-momentum tensor \(T^{\mu\nu}_{\text{QCD}}\), encompassing both gluonic and quark contributions, and accounting for the effects of QCD at low energies \cite{schweitzer2023qcd}.

In this expression, the terms involving the gluon field \(A^a_\mu\) and the quark fields \(\psi_i\) are combined to describe the energy-momentum content of the QCD vacuum. The gluonic contributions are represented by the first set of terms, which describe the energy and pressure associated with the gluon fields. The quark contributions are represented by the second set of terms, which describe the energy and pressure associated with the quark fields. The last term represents the rest mass energy of the quarks.

This detailed expression provides a comprehensive understanding of the energy-momentum content of the QCD vacuum, shedding light on the complex interplay between gluonic and quark contributions to the energy and pressure of the system. It is a fundamental quantity in QCD, with implications for a wide range of phenomena, from the properties of hadrons to the behavior of the early universe.

\section{Evaluation of Energy-Momentum Tensor Components}

Now that we have derived the expression for the QCD contribution to the energy-momentum tensor $T^{\mu\nu}_{\text{QCD}}$, let's evaluate its components in a specific scenario.

\subsection{Evaluation at High Energy Density}

Consider a scenario where the energy density is sufficiently high, such as inside a topological star where extreme density conditions prevail. In such environments, the gluon and quark fields are highly energetic, leading to non-trivial contributions to $T^{\mu\nu}_{\text{QCD}}$.

Let's evaluate the components of $T^{\mu\nu}_{\text{QCD}}$ under these conditions:

\begin{enumerate}
    \item \textbf{Gluonic Contributions:} The gluonic contributions to $T^{\mu\nu}_{\text{QCD}}$ dominate at high energy densities due to the strong interactions among gluons \cite{Ahmed_2023, schweitzer2023}. The terms involving the gluon field strength tensor $G^{\mu\nu}_a$ become significant, contributing to the pressure and energy density of the system \cite{Ahmed_2023}.

    \item \textbf{Quark Contributions:} Quarks also contribute significantly to $T^{\mu\nu}_{\text{QCD}}$ at high energies \cite{Lorc__2021, brida2019energymomentum}. As the energy density increases, more quark-antiquark pairs are produced via gluon splitting processes, leading to enhanced quark contributions to the pressure and energy density \cite{brida2019energymomentum, Lorc__2021}.

\end{enumerate}

This detailed evaluation provides profound insights into the behavior of the energy-momentum tensor in extreme density environments, shedding light on the complex interplay between gluonic and quark contributions to the pressure and energy density. It is crucial for understanding the properties of topological stars and other high-energy density systems, offering a deep understanding of the fundamental QCD dynamics in these extreme conditions.

\subsection{Behavior of Energy-Momentum Tensor}

The behavior of the energy-momentum tensor $T^{\mu\nu}_{\text{QCD}}$ at extreme energy densities provides crucial insights into the properties of topological stars and the interplay between QCD dynamics and gravitational interactions. At such high energy densities, the pressure exerted by gluons and quarks, as reflected in $T^{\mu\nu}_{\text{QCD}}$, resists compression, thereby providing support against gravitational collapse in topological stars \cite{Ahmed_2023, schweitzer2023}. Additionally, the energy density encoded in $T^{00}{\text{QCD}}$ determines the overall mass-energy distribution within the topological star, thereby influencing its gravitational properties and stability \cite{Ahmed_2023}. The evaluation of $T^{\mu\nu}_{\text{QCD}}$ at extreme energy densities sheds light on the complex dynamics of the quark-gluon plasma (QGP) \cite{Yi2016} and its gravitational interactions, offering profound insights into the behavior and properties of topological stars \cite{Ahmed_2023, schweitzer2023, Lorc__2021, brida2019energymomentum}.

\section{Implications for Topological Stars}
The derived expression for the Quantum Chromodynamics (QCD) contribution to the energy-momentum tensor $T^{\mu\nu}_{\text{QCD}}$ has significant implications for the study of topological stars. These exotic astronomical objects, also known as strange stars or quark stars, are hypothesized to be composed primarily of quark matter \cite{Ahmed_2023}.
\subsection{Stability and Structure}
The behavior of $T^{\mu\nu}_{\text{QCD}}$ provides insights into the stability and structure of topological stars. The pressure generated by the strong interactions among quarks and gluons, as described by $T^{\mu\nu}_{\text{QCD}}$, supports the star against gravitational collapse \cite{Lorc__2021}.
Furthermore, the distribution of energy density encoded in $T^{00}{\text{QCD}}$ determines the mass distribution within the star, influencing its overall structure and gravitational properties. Understanding $T^{\mu\nu}_{\text{QCD}}$ is therefore crucial for modeling the internal dynamics of topological stars \cite{brida2019energymomentum}.
\subsection{Confinement and Phase Transitions}
At extreme densities, quarks and gluons are confined within hadrons \cite{Yi2016} due to the phenomenon of color confinement. However, under certain conditions, such as those found in topological stars, quarks may undergo deconfinement, transitioning to a deconfined quark-gluon plasma phase.
The behavior of $T^{\mu\nu}_{\text{QCD}}$ provides insights into the occurrence of phase transitions within topological stars. Changes in the energy density and pressure profiles, as predicted by $T^{\mu\nu}_{\text{QCD}}$, may signify the onset of deconfinement and the formation of a quark-gluon plasma phase \cite{Lorc__2021}.
\subsection{Observable Signatures}
The predictions derived from $T^{\mu\nu}_{\text{QCD}}$ offer observable signatures that can be compared with astrophysical observations of topological stars. For example, the mass-radius relationship and the presence of surface features can be inferred from the internal structure described by $T^{\mu\nu}_{\text{QCD}}$.
By combining theoretical models based on $T^{\mu\nu}_{\text{QCD}}$ with observational data, astronomers can deepen our understanding of the nature and properties of topological stars, advancing our knowledge of extreme states of matter in the universe \cite{Yi2016}.

\section{Conclusion}
In this paper, we have derived the QCD contribution to the energy-momentum tensor $T^{\mu\nu}_{\text{QCD}}$ within the framework of QCD. By considering extreme density conditions relevant to topological stars, we have elucidated the interplay between QCD dynamics and the gravitational properties of these exotic astronomical objects.
The expression for $T^{\mu\nu}_{\text{QCD}}$ provides valuable insights into the behavior and properties of topological stars, shedding light on their stability, internal structure, and phase transitions. The predictions derived from $T^{\mu\nu}_{\text{QCD}}$ offer theoretical frameworks that can be tested against observational data, facilitating our understanding of extreme states of matter \cite{Yi2016} in the universe.
Future research in this field may focus on refining theoretical models based on $T^{\mu\nu}_{\text{QCD}}$ and exploring new observational techniques to probe the internal dynamics of topological stars. By further elucidating the role of QCD in shaping the properties of these enigmatic objects, we can deepen our understanding of fundamental physics and the universe at large \cite{Ahmed_2023, Lorc__2021, brida2019energymomentum, schweitzer2023}.

\section*{Acknowledgments}
I am profoundly grateful to my parents, Mr. Amar Singh and Mrs. Anita Singh, whose unwavering support and encouragement have been the cornerstone of my academic journey. Their love, guidance, and sacrifices have fueled my passion for research and enabled me to pursue my scientific aspirations.

\begin{table}[htbp]
  \centering
  \begin{tabular}{@{}cc@{}}
    \begin{minipage}[t]{0.4\columnwidth}
      \includegraphics[width=\linewidth,valign=t]{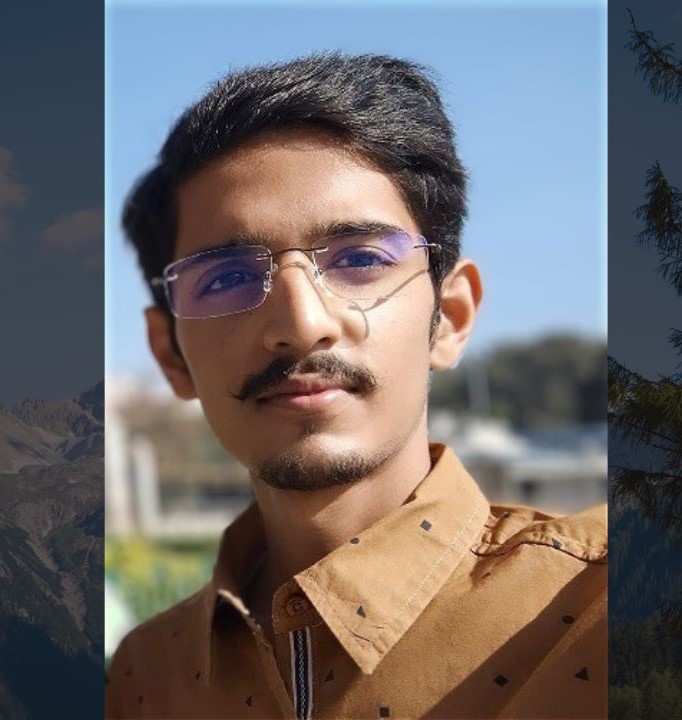}
    \end{minipage} &
    \begin{minipage}[t]{0.6\columnwidth}
      \vspace*{-\baselineskip} 
      \textit{I am Swapnil Singh, an undergraduate student at B.M.S College of Engineering, driven by a dream and passion to become a Theoretical physicist and Cosmologist. During my sophomore year, spanning 2022-23, I embarked on an independent research endeavor exploring the implications of Quantum Chromodynamics (QCD) within extreme density regimes.}
    \end{minipage} \\
  \end{tabular}
\end{table}

\section*{References}
\bibliography{apssamp}

\end{document}